\documentclass[aip,rsi, preprint,amsmath, amssymb]{revtex4-2}

\usepackage{graphicx}
\usepackage{dcolumn}
\usepackage{bm}

\usepackage[utf8]{inputenc}
\usepackage[T1]{fontenc}
\usepackage{mathptmx}
\usepackage{etoolbox}
\usepackage{amsmath}

\usepackage{color}

\makeatletter
\def\@email#1#2{%
 \endgroup
 \patchcmd{\titleblock@produce}
  {\frontmatter@RRAPformat}
  {\frontmatter@RRAPformat{\produce@RRAP{*#1\href{mailto:#2}{#2}}}\frontmatter@RRAPformat}
  {}{}
}%
\makeatother
\begin{document}


\title{Photoelectron angular distribution as a linear polarization analyzer for soft and tender X-rays}
\author{Yoshiyuki Ohtsubo}
\email{y\_oh@qst.go.jp}
\affiliation{NanoTerasu Center, National Institutes for Quantum Science and Technology (QST), Sendai, Miyagi, 980-8579, Japan}
\author{Hiroaki Kimura}%
\affiliation{NanoTerasu Center, National Institutes for Quantum Science and Technology (QST), Sendai, Miyagi, 980-8579, Japan}

\date{\today}

\begin{abstract}
Although the polarization of soft and tender X-rays is widely used to investigate the diverse physical properties of materials, experimental methods to determine the polarization of tender X-rays (1.5–3.0 keV) remain limited.
To address this issue, we propose a method  based on the photoelectron angular distribution to detect the polarization of X-rays in this energy range.
The angular distribution of photoelectrons emitted from carbon targets was measured using linearly polarized 0.2 to 3.0 keV synchrotron radiation.
The photoelectron intensity depends on the angle between the photon's electric field vector and the direction at which the photoelectron is emitted from the target.
This result indicates that the photoelectron angular distribution can be used to reliably determine the linear polarization of soft and tender X-rays over a wide range of energy.
\end{abstract}

\maketitle

\section{Introduction}
Experimental methods using polarized photons provide a wide range of physical information, from biomolecular dynamics \cite{Bio-UV} to astronomical phenomena such as gamma-ray bursts \cite{Astro-HX}.
Measurements using polarized X-rays are known as powerful element-specific tools that reveal the electronic structure of materials, such as spin and orbital magnetic moments by X-ray magnetic circular and linear dichroism \cite{XMCD, Vaz2025}, the ground-state symmetry of correlated systems by photoelectron spectroscopy \cite{Fujiwara16}, and selective electronic and magnetic excitations by resonant inelastic X-ray spectroscopy \cite{Braicovich14}.
For such methods, a crucial initial step is to determine the polarization of the photons involved in the experiments, including the incident beam, fluorescence emission, and inelastic scattering.

For X-rays, Bragg reflection from periodic structures at a reflection angle close to  Brewster's angle is widely used as an X-ray polarization analyzer.
Although periodic stackings of multilayer thin films are suitable reflecting targets for soft X-rays \cite{Schafers99}, the maximum photon energy for multilayer Bragg reflectors is approximately 1.2 keV because, beyond this energy, the required multilayer period becomes too short \cite{Wang2011}.
In contrast, the periodic atomic structures of single crystals can serve as Bragg reflectors for hard X-rays \cite{Bergevin95, Detlefs2012}.
The lower-energy limit for crystal-based Bragg reflectors depends on the lattice constant and is typically about 3 keV for diamond crystals.
Extensive studies have been done to expand the photon energy range for polarization measurements; for example,
the use of Si crystals instead of diamond has lowered the energy threshold to around 2 keV \cite{Bouchenoire2012}, and combining multilayer phase retarders with beryl crystal analyzers enables complete (both linear and circular) polarization measurements from 1.1 to 1.5 keV \cite{Wang12}.
Unfortunately, the energy window for each method is limited due to the strict correspondence required between the periodic length and the photon energy, so measuring photon polarization over a wide energy range requires the frequent change of reflectors.
Thus, despite the rich physical information provided by polarization measurements, as exampled by the above-mentioned experimental methods, polarimetry methods for the intermediate-energy region between soft and hard X-rays (i.e., tender X-rays; 1.5–3.0 keV), remain severely limited.

To address this issue, we revisit the fundamental principles of photoemission and focus on the angular distribution of photoelectrons excited by linearly polarized X-rays for probing the incident photon polarization.
The validity of this polarimetry method was demonstrated across a wide energy range from 200 to 3000 eV by measuring the photoelectron angular distribution from a reference carbon target using a rotating-analyzer method.
We then use a simple model to qualitatively explain the polarization dependence of the angular distribution of photoelectrons.

\section{Methods}
Soft and tender X-rays were provided by NanoTerasu BL13U \cite{Ohtsubo_2025}.
An APPLE-II undulator generated linearly polarized synchrotron radiation, and the beam was monochromatized for the polarization measurements by using a varied-line-spacing plane grating.
Both linear horizontal (LH) and linear vertical (LV) polarizations were used.
While this beamline has a segment of four APPLE-II undulators\cite{Inaba26}, single undulator segment was used for most of the measurements in this work, except where noted.
The resolving power ($E/dE$) of the incident beam was 3000, 1400, and 1000 for 0.4, 2.0, and 3.0 keV photons, respectively, unless otherwise specified.
$E/dE$ decreases with increasing photon energy.
Polarization was measured by using a polarimetry apparatus based on a rotating analyzer method \cite{Kimura04}.

\begin{figure}
\includegraphics{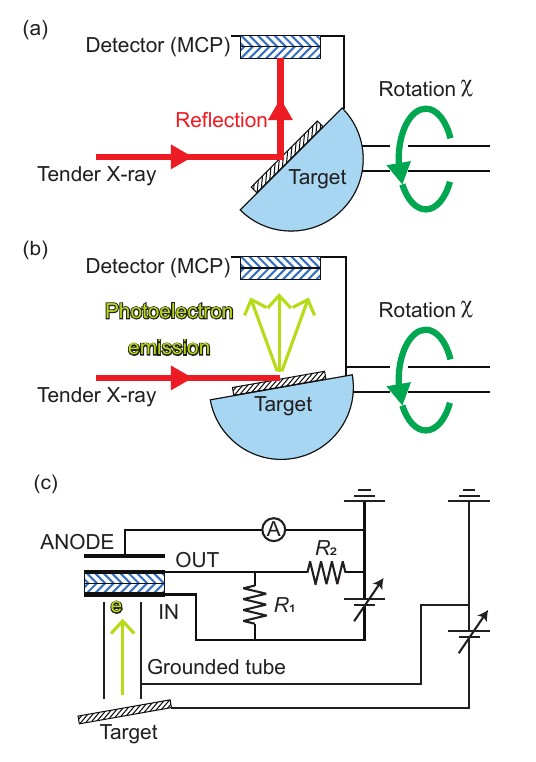}%
\caption{\label{fig1} 
(a) Schematic drawing of the polarization measurement using a multilayer reflection target and a rotating analyzer.
(b) The same as panel (a), but using photoelectrons emitted from the target. 
In both cases, the target and the MCP detector rotate around the axis of the incident beam, and the MCP signals are recorded as a function of the rotation angle $\chi$.
(c) Circuit diagram for the MCP detector. The resistances are $R_1$ = 2 M$\Omega$ and $R_2$ = 0.1 M$\Omega$.
}%
\end{figure}

Figure \ref{fig1} shows schematically the experimental geometry of the incident beam, target, and detector.
Figure \ref{fig1}(a) shows the configuration for the rotating-analyzer method using a multilayer soft-X-ray Bragg reflector \cite{Kimura04}.
The incident beam is reflected by an analyzer target (e.g., a Cr/C multilayer) and subsequently detected by a microchannel plate (MCP) detector (Hamamatsu Photonics, F-4655).
At the pseudo-Brewster angle, the analyzer reflects the incident beam when its electric field is perpendicular to the plane of incidence ($s$-polarization geometry).
In contrast, the reflectivity approaches zero when the electric field lies within the plane of incidence ($p$-polarization geometry).
By rotating the azimuthal angle $\chi$ of the target together with the detector, the major axis of the polarization ellipse (polarization azimuth) and the degree of linear polarization of the incident beam can be determined.
In this work, the incident beam is always in the horizontal plane, and $\chi=0^\circ$ is defined as the beam being reflected vertically at a $45^{\circ}$ angle of incidence with respect to the multilayer surface.

As shown in Fig. \ref{fig1}(b), the photoelectron angular distributions were measured using the same apparatus as for the rotation analyzer measurements described above.
The angle of incidence of the target was set to $4\pm1^\circ$ for a high photoelectron yield in the soft and tender regions of the X-ray energy.
Photoelectrons emitted from the target were detected by the MCP detector.
A stacked graphite sheet (Panasonic Graphite TIM) and a glassy carbon plate (SPI Supplies) were used as carbon targets.
In addition, a single-crystal Si(111) substrate and a polycrystalline Cr thin film ($\approx$200 nm thick) also served as target samples.

Figure \ref{fig1}(c) shows the circuit diagram of the MCP detector.
A single power supply provides negative voltages to both the entrance (IN) and exit (OUT) of the MCP, and the multiplied electrons are subsequently detected as an anode current ($I_\text{MCP}$).
The voltages at IN and OUT are determined by the ratio $R_1/R_2$ of the resistances.
The applied bias was $1.4\pm0.1$ kV, which also serves as a blocking voltage against low-energy electrons.
Although $I_\text{MCP}$ obtained in the grazing-incidence geometry [Fig. \ref{fig1}(b)] were significantly weaker than the X-rays reflected by the multilayer targets, increasing the MCP voltage from $\approx$1.0 kV to $\approx$1.4 kV (corresponding to a gain of $\times$2400) resulted in sufficient anode current ($I_\text{MCP}\approx100$ pA).
In this work, unless otherwise specified, the maximum $I_\text{MCP}$ for a single $\chi$ scan ($I_\text{max}$) was tuned to 10--150 pA.
In this range, $I_\text{max}$ had no influence on the angular-distribution measurements.
A separate power supply was used to apply the sample bias.
Additionally, a grounded metal tube positioned between the target and the MCP restricts the detectable solid angle to $\pm2^{\circ}$.

\section{Results and discussion}
\begin{figure}
\includegraphics{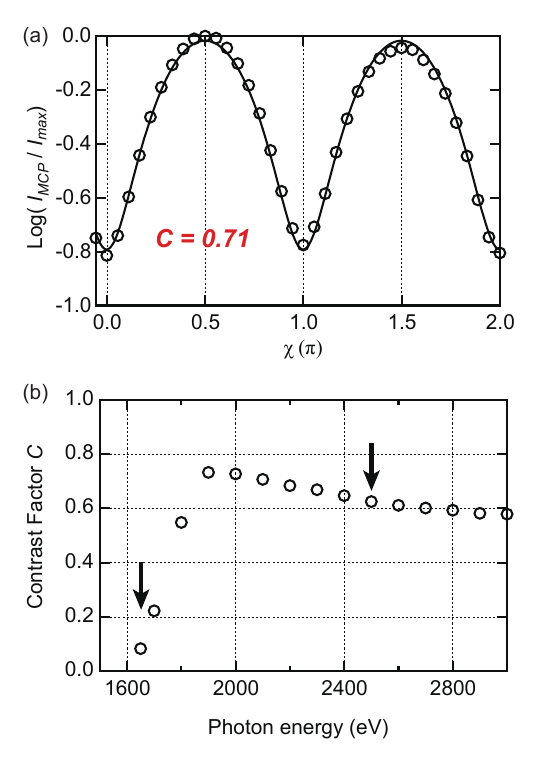}%
\caption{\label{fig2}
(a) MCP anode current $I_\text{MCP}$ as a function of $\chi$ (circles) measured using tender X-ray synchrotron radiation ($h\nu$ = 2000 eV).
The incident beam is polarized along the linear horizontal orientation.
The solid curve is the fit (see main text for details), which gives the contrast factor $C=0.71$.
(b) Contrast factors obtained from $I_\text{MCP}$ with $\chi$ rotation as a function of incident photon energy with linear polarization.
Arrows indicate the data shown in Fig. \ref{fig3}.
}%
\end{figure}

Figure \ref{fig2}(a) shows the normalized $I_\text{MCP}$ from the graphite target as a function of $\chi$ for incident tender X-rays with LH polarization ($h\nu$ = 2000 eV).
$I_\text{MCP}$ varies periodically with $\chi$, reaching a maximum and minimum at $(n+1/2)\pi$ and $n \pi$ ($n \in \mathbb{Z}$), respectively.
This modulation is consistent with the incident photon polarization.

The $\chi$-dependent intensity detected by using a rotating analyzer can be described by
\begin{eqnarray}
I_\text{MCP}(\chi) = I_0 \left( \frac{2C}{1+C} \sin^2(\chi-\alpha) + \frac{1-C}{1+C} \right),
\label{eq:ELLI}
\end{eqnarray}
where $I_0$ is a scaling factor, $\alpha$ is the electric-field angle of the linearly polarized incident beam, and $C$ is the contrast factor (0 $\leq C \leq$ 1).
$\alpha$ = $n \pi$ and $(n+1/2)\pi$ correspond to the LH and LV polarization of the incident beam, respectively ($n \in \mathbb{Z}$).
In the ideal case (i.e., perfect linear polarization and a perfect analyzer without misalignment) $C=1$ and $I_0$ corresponds to the maximum observed intensity ($I_\text{max}$).
In contrast, for an isotropic angular distribution of $I_\text{MCP}$ that is expected for the unpolarized X-rays, $C$ becomes 0.
$I_0=I_\text{max}$ if the polarimeter is perfectly aligned, in which case $I_\text{MCP}(\chi) = I_\text{MCP}(\chi + \pi)$ because a $\pi$ rotation in $\chi$ is ideally equivalent due to the rotational symmetry of the detection system about the beam axis.
The solid line in Fig. \ref{fig2}(a) is the fit based on Eq. (\ref{eq:ELLI}), with $C$, $I_0$, and $\alpha$ as free parameters.
The values obtained for $C$ and $\alpha$ are 0.71 and 0.00, respectively.
As shown in the figure, the fit faithfully reproduces the experimental data.

Similar values for the contrast factor and $\alpha$ were obtained using the glassy carbon target.
Notably, Eq. (\ref{eq:ELLI}) is mathematically equivalent to the $\chi$-dependence detected by using the multilayer reflectors \cite{Kimura04}.
The primary difference between the Bragg reflection and the results in Fig. \ref{fig2}(a) is a $\pi$/2 phase shift in $\alpha$.
Specifically, $s$-polarized ($p$-polarized) incidence results in a maximum (minimum) Bragg reflection, whereas it corresponds to minimum (maximum) $I_\text{MCP}$ in the current setup.
In addition, the influence of $E/dE$ on the measurements was checked at 2.1 keV, yielding $C=0.72\pm0.01$ for $E/dE$ in the range 1000--5000, suggesting that an energy variation of a few eV in the incident beam plays a minor role in this measurement.

Figure \ref{fig2}(b) shows the contrast factor $C$ obtained from the fitting procedure described above at various incident photon energies.
For the entire energy range shown in Fig. \ref{fig2}(b), except for the notably small $C$ at 1650 eV, Eq. (\ref{eq:ELLI}) agrees well with the measured spectra.
The fitting parameter $\alpha$ is approximately zero ($\pm0.5\pi$) for the LH (LV) polarization, with deviations smaller than $0.01\pi$, which correctly reflects the polarization of the incident beam.
Furthermore, the difference between $I_0$ and $I_\text{max}$ remained small in most cases.
These results indicate that, in the geometry depicted in Fig. \ref{fig1}(b), the carbon target functions as a polarization analyzer, analogous to the Bragg reflection polarizer used for reflected photons.

Note that the degree of linear polarization of the incident beam exceeds 0.99, as is well established for X-ray synchrotron radiation from an APPLE-II undulator \cite{pol_APPLE}.
Therefore, the contrast factor obtained in this work is nearly equivalent to the linear polarization selectivity.
This factor represents the performance of the linear polarization analyzer, corresponding to the polarizance of Bragg reflection targets.
As shown in Fig. \ref{fig2}(b), the contrast factor reaches its maximum between 1800 and 2000 eV and then gradually decreases at higher photon energies; however, it remains above 0.5 even at 3000 eV.
This result indicates that the carbon target is an effective polarizer for measuring incident photon polarization across this broad energy range, in contrast with Bragg reflectors, which are limited by a narrow energy window determined by their periodic length.

\begin{figure}
\includegraphics{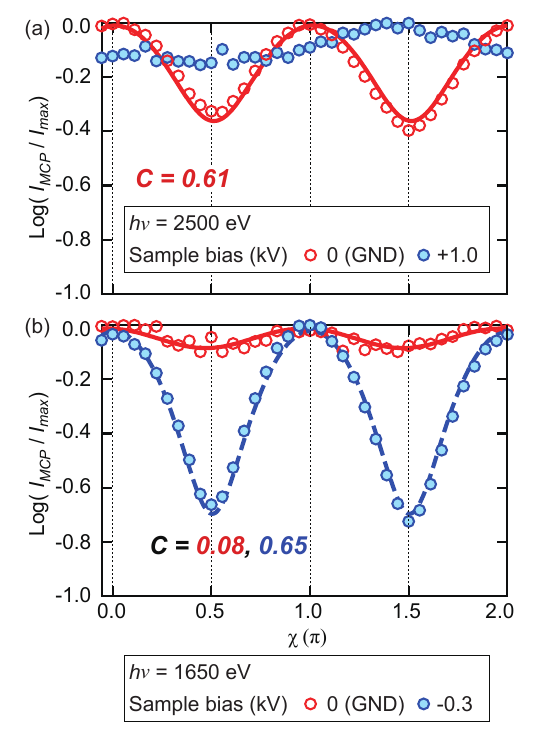}%
\caption{\label{fig3} 
MCP anode current ($I_\text{MCP}$) as a function of $\chi$ measured with (open circles) and without (filled circles) applying a sample bias voltage.
The incident beam is polarized along the linear vertical orientation.
Solid and dashed curves represent the fit obtained using Eq. (\ref{eq:ELLI}). Photon energy is (a) 2500 eV and (b) 1650 eV.
}%
\end{figure}

A remaining question regarding the carbon ``polarization analyzer'' is the physical origin of the $\chi$-depdendent $I_\text{MCP}$.
MCPs are sensitive to both photons and electrons, both of which are detected in the current setup.
To distinguish between these two contributions, a sample bias was applied at $h\nu$ = 2500 eV and 1650 eV, as shown in Figs.~\ref{fig3}(a) and \ref{fig3}(b), respectively.
At 2500 eV [Fig. \ref{fig3}(a)], the contrast factor of 0.61 obtained with the grounded target vanishes when a positive sample bias of +1.0 kV is applied, reducing the kinetic energy of the photoelectrons.
Under this bias, the measured spectrum is no longer reproduced by Eq. (\ref{eq:ELLI}).
Furthermore, $I_\text{MCP}$ itself drops by nearly two orders of magnitude as the sample bias increases from 0 (grounded) to +1.0 kV.
Together with the case of photoelectron acceleration shown in Fig. \ref{fig3}(b), these results strongly suggest that the $\chi$-dependent $I_\text{MCP}$ detected in this work originates primarily from photoelectrons.

The improved contrast factor $C$ at 1650 eV achieved with an acceleration voltage [Fig. \ref{fig3}(b)], contrasted with the decay of $C$ below 1800 eV observed in Fig. \ref{fig2}(b), suggests a detection threshold for photoelectron kinetic energy at approximately 1700 eV in the current experimental setup.
As shown in Fig. \ref{fig1}(c), the bias applied to the MCP (1.3--1.4 kV) acts as a blocking voltage.
Without the sample bias, photoelectrons generated by 1650 eV photons would reach the MCP with at most $\approx$300 eV of kinetic energy.
Furthermore, because the photoelectron path is enclosed by a grounded metal tube, most of these low-energy electrons are likely attracted to and absorbed by the tube walls before reaching the detector.
These geometric and electrostatic effects likely account for the observed threshold at $\approx$1700 eV.

\begin{figure}
\includegraphics{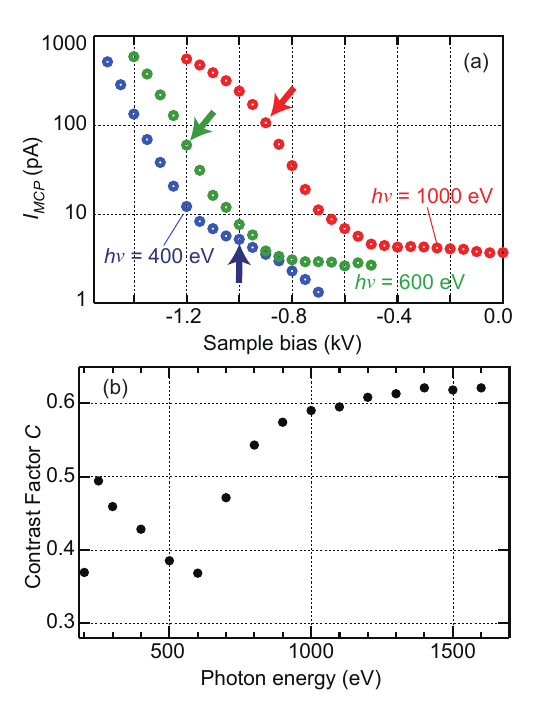}%
\caption{\label{fig4} 
(a) $I_\text{MCP}$ as a function of sample bias for $h\nu$ = 400, 600, and 1000 eV in the $p$-polarized geometry. Arrows indicate the sample bias where the contrast factor is maximal.
(b) Maximum contrast factors as a function of incident photon energy, measured with sample biases ranging from $-1.2$ to $-0.3$ kV.
}%
\end{figure}

Figure \ref{fig4}(a) shows $I_\text{MCP}$ as a function of sample bias measured with various photon energies to investigate the effect of bias voltage.
In this measurement, the four APPLE-II undulator segments were used together to obtain high flux at in low-sample-bias region.
The odd (1, 3) and even (2, 4) segments were set to LV and LH polarizations, respectively, and their interference condition was tuned to yield a composite linear polarization with $\chi = \pi/4$ \cite{Inaba26}.
Figure \ref{fig4}(b) is the maximum contrast factors as a function of incident photon energy.
For this measurement, single APPLE-II undulator segment was used again, to ensure the purity of the incident-beam linear polarization above 0.99.
At $h\nu$ = 1000 eV, $I_\text{MCP}$ exhibits a small rise around $-0.6$ kV, increases exponentially beyond $-0.8$ kV, and saturates beyond approximately $-1.0$ kV.
This three-step behavior can be understood by considering the detector setup as a retarding-field analyzer (RFA) of photoelectron kinetic energy.
In this scenario, the first edge corresponds to valence electrons in the carbon target (binding energy $E_b$ of a few eV) that are accelerated by the sample bias to reach the MCP anode against the blocking voltage (fixed at 1.3 kV for the series of measurements shown in Fig. \ref{fig4}).
The second exponential increase beyond $-0.8$ kV is assigned to the carbon $1s$ core level ($E_b \approx 285$ eV).
At more negative biases down to $-1.2$ kV, secondary electrons further enhance the $I_\text{MCP}$ signal.
The contrast factor reaches its maximum around a sample bias of $-0.9$ kV for $h\nu$ = 1000 eV, suggesting that carbon $1s$ photoelectrons play a major role for large contrast factors.

At photon energies $h\nu$ = 600 and 400 eV, $I_\text{MCP}$ increased exponentially beyond a threshold sample bias, similar to the case of $h\nu$ = 1000 eV.
However, the sample bias at which the contrast factor is maximal changes: for 600 and 1000 eV, it occurs in the middle of the exponential rising edge, whereas for 400 eV, it shifts to around $-1.0$ kV; this is prior to the inflection point of $I_\text{MCP}$ (near $-1.2$ kV).
This variation also appears in Fig. \ref{fig4}(b), which plots the maximum contrast factor as a function of the incident photon energy.
In the 900--1500 eV range, the maximum contrast factor stays almost constant around 0.6, slightly decaying with decreasing photon energy.
In the 600--900 eV range, the contrast factor decays to a minimum of 0.38 at 600 eV but then rises again in the 300--600 eV range.
This behavior suggests that the fractional contribution to the photoelectron angular distribution changes depending on the incident photon energy, as discussed later.
Nevertheless, from a practical point of view, the photoelectron angular distribution from the biased carbon target can serve as a photon polarimeter in this energy range.

To gain further insight into how $I_\text{MCP}$ depends on incident photon energy, $\chi$, and sample bias, we used a simple isolated atom model to calculate the angular distribution of the photoelectrons \cite{Cooper93, Trzhaskovskaya01}.
The angular distribution of photoelectrons excited by linearly polarized photons is described by
\begin{eqnarray}
J(\theta, \phi) = \frac{d\sigma_i}{d\Omega} = \frac{\sigma_i}{4\pi}[1+\beta P_2({\rm cos}\theta)+(\delta + \gamma{\rm cos}^2\theta){\rm sin}\theta{\rm cos}\phi]
\label{eq:two},
\end{eqnarray}
where $\sigma_i$ is the subshell photoionization cross section, and $P_2({\rm cos}\theta)$ = $\frac{1}{2}(3{\rm cos}^2\theta-1)$ is the second-order Legendre polynomial.
The parameter $\beta$ is the dipole angular distribution parameter, and $\gamma$ and $\delta$ are the nondipole angular distribution parameters.
$\theta$ is the angle between the electric field ${\bm \epsilon}$ of the (linearly polarized) incident photons and the photoelectron momentum ${\bm p}$, and $\phi$ is the angle between the photon-propagation axis and the plane defined by ${\bm \epsilon}$ and ${\bm p}$.
The parameters $\sigma_i, \beta, \gamma,$ and $\delta$ are tabulated in Ref. \onlinecite{Trzhaskovskaya01} for discrete photon energies.
In this work, we obtained the values of $\beta, \gamma,$ and $\delta$ by interpolating between the tabulated data.
For $\sigma_i$, the interpolation was done on a logarithmic scale, accounting for its exponential decay with increasing photon energy; $h\nu$ = photoelectron energy + binding energy + work function.
For linearly polarized photons, ${\bm \epsilon}$ is always perpendicular to the incident photon axis.
In the $p$-polarization ($s$-polarization) geometry, ${\bm p}$ is parallel (perpendicular) to ${\bm \epsilon}$, such that $\theta$ = 0 ($\pi$/2).
The angle $\phi$ is $\pi$/2 for $s$ polarization but is undefined for $p$ polarization.
Note that this indefiniteness poses no problem because the ${\rm cos}\phi$ term in Eq. (\ref{eq:two}) vanishes regardless.

Before going into the quantitative details, let us examine the simplest case of Eq. (\ref{eq:two}, where $\sigma_i = 1$, $\beta = 2$, and $\delta = \gamma = 0$.
This represents that the angular distribution $J$ is purely governed by the electric dipole transition from a $s$ atomic orbital as an initial state.
In this case, the azimuthal angle $\phi$ is irrelevant because the $\cos\phi$ term vanishes, simplifying the equation to $J(\theta) \propto \cos^2\theta$.
Here, $\theta$ is equivalent to the experimental parameter $\chi$.
The $p$- and $s$-polarization geometry are realized with $\chi-\alpha = (n+1/2)\pi$ and $n\pi$, respectively; for example, both the cases of LH ($\alpha$ = 0) with $\chi$ = $\pi$/2 and LV ($\alpha = \pi/2$) with $\chi$ = 0 correspond to $\theta$ = 0.
This behavior is identical to eq. (\ref{eq:ELLI}) with the contrast factor $C$ = 1.
Therefore, in this simplest case governed by the electric dipole transition from a $s$ atomic orbital, the photoelectron angular distribution qualitatively agrees with the experimental observations.
For quantitative consideration, numerical calculations are examined in the following paragraphs.

\begin{figure}
\includegraphics{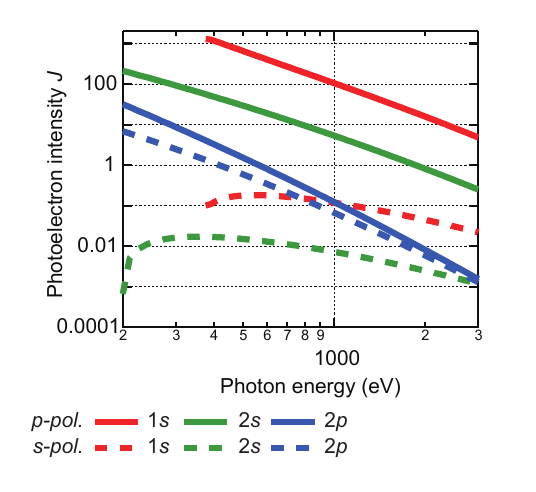}%
\caption{\label{fig5}
Photoelectron distributions calculated as a function of the incident photon energy from Eq. (\ref{eq:two}) for $p$- and $s$-polarization geometries.
To simplify the calculation, the solid angle of the MCP detector was not considered.
}%
\end{figure}

Based on Eq. (\ref{eq:two}), Fig. \ref{fig5} shows that the photoelectron distributions depend on the incident photon energy for both $p$- and $s$-polarization geometries, namely $J(\theta = 0, \phi: undefined)$ and $J(\theta = \pi/2, \phi = \pi/2)$, respectively.
For the carbon $1s$ and $2s$ initial states, $J_p$ is much larger than $J_s$, where $J_p$ ($J_s$) corresponds to the intensity $J(\theta, \phi)$ calculated for the $p$-polarization ($s$-polarization) geometry.
In contrast, the difference between $J_p$ and $J_s$ is less than an order of magnitude for $2p$ states.
These results can be explained according to the simplest model discussed above.
In the case of carbon $1s$ and $2s$, photoelectron emission process is predominantly governed by the electric dipole excitation from the $s$ orbitals ($\beta \simeq$ 2, $\delta$ and $\gamma$ are very small) \cite{Trzhaskovskaya01}.
For carbon $2p$, $\beta$ is smaller than 2 and depends on the photon energy, because the $2p$ initial state is excited differently from $s$ orbitals.
In addition, the photoelectrons from $1s$ is always dominant against the valence electrons ($2s$ and $2p$) as far as it can be excited with the photon energy higher than $E_b$.

The dominant contribution of $J_p$ over the entire energy range qualitatively explains why the $p$-polarization geometry produces a larger $I_\text{MCP}$ in the experiments.
However, from a quantitative viewpoint, if only the $J_p$ and $J_s$ values calculated from Eq. (\ref{eq:two}) contributed to the experiment, the contrast factor should exceed 0.99.
To explain this discrepancy, other sources of $I_\text{MCP}$ must be taken into account.
First, the secondary electrons follow a nearly isotropic angular distribution because the anisotropic distribution is averaged out through multiple scatterings before emission.
Second, the MCP also detects fluorescent photons, whose polarization dependence differs from that of elastic photoelectrons, thereby reducing the overall contrast factor.
Third, the KLL Auger electrons can also significantly reduce the contrast factor at lower photon energies, since such an Auger emission has an isotropic angular distribution. The kinetic energy of the C KLL Auger electron is 260-280 eV\cite{Moulder95}.

Based on the possible $I_\text{MCP}$ sources discussed above, we now discuss the general trend of the contrast factors observed in Figs. \ref{fig2}(b) and \ref{fig4}(b).
At photon energies above 1000 eV, Auger electrons are almost negligible, because they cannot reach the MCP detector without a sample bias.
Even with a bias, the elastic photoelectrons reach the detector at a lower bias voltage [see the case of $h\nu$ = 1000 eV in Fig. \ref{fig4}(a)].
In this case, elastic electrons provide an anisotropic distribution based on the electric dipole transition, together with the isotropic background from secondary electrons and fluorescent photons, resulting in contrast factors of 0.6--0.7.
As the incident photon energy increases, the photoelectron energy window detectable in our retarding-field-like setup expands, thereby integrating a larger contribution of secondary electrons.
It likely explains the gradual decay of the contrast factor observed in Fig. \ref{fig2}(b).
For soft X-rays with $h\nu$ = 600--900 eV, the contribution of the isotropic Auger electrons becomes larger, causing the contrast factor to decay.
Then, the contrast factor increases again below 500 eV.
In this region, the maximum contrast factor occurs before the inflection point of the $I_\text{MCP}$ curve (e.g., at $-1.0$ kV for $h\nu=400$ eV).
Here, the main source of photoelectrons is the elastic electrons from the carbon $2s$ and $2p$ valence bands, which are also excited by the electric dipole transition (as shown in Fig. \ref{fig5}), thereby restoring the anisotropic distribution.
Sample biases corresponding to the valence electrons showed a small contrast factor at higher photon energies (such as around $-0.6$ kV for $h\nu$ = 1000 eV) due to insufficient flux of elastic $2s$ and $2p$ photoelectrons.
This result stems from the exponential decay of the photoelectron intensity (see Fig. \ref{fig5}).
The only feature that is not explained in the above discussion is the small contrast factor of 0.37 at $h\nu=200$ eV.
It is difficult to conclude the reason for this small value because of the limited electron energy resolution and low values of $I_\text{MCP}$.
From the practical viewpoint, let us note that the photon polarimetry measurement itself is still available at $h\nu$ = 200 eV in the current experimental setup, the rotating analyzer with the MCP detector.

The discussion above suggests that the contrast factor could be improved by detecting monochromatic, elastic photoelectrons without fluorescent photons, for example, by using an electron energy analyzer without electron-path distortion.
It should be noted that the present setup has a non-negligible difference from the ideal retarding-field analyzer, that is, the grounded metal tube lying on the photoelectron path.
Because it is always grounded irrespective to the sample bias, its magnification parameter as electron lens varies with the sample bias and thus the photoelectron acceptance angle changes at the same time.
Nevertheless, our results demonstrate that the MCP detector in a rotating-analyzer geometry is practically sufficient to determine the electric field orientation (i.e., the major axis of the polarization ellipse).
Furthermore, the degree of linear polarization can be determined by calibrating the target's polarization selectivity (contrast factor) at the relevant photon energy using a standard beam, such as X-ray synchrotron radiation from an APPLE-II undulator.

\begin{table}
\caption{\label{tab1} Contrast factors from the photoelectron angular distributions for each target element at $h\nu$ = 2000 eV.}
\begin{ruledtabular}
\begin{tabular}{cccc}
&Carbon&Silicon&Chromium\\
\hline
Experiment&0.71&0.14&0.38\\
\end{tabular}
\end{ruledtabular}
\end{table}

From a practical point of view, two different materials were also examined as a photoelectron target for tender X-ray polarimetry.
Table \ref{tab1} summarizes the photoelectron contrast factors obtained using carbon, silicon, and chromium targets.
The results show that carbon significantly outperforms the heavier elements Si and Cr.
Regarding the target dependence on the polarimetry, the crystal structure does not play a major role in determining the photoelectron contrast; both the two-dimensional layered graphite and the three-dimensional amorphous glassy carbon yielded nearly identical experimental results, utilizing the carbon $1s$ photoemission.

\begin{table}
\caption{\label{tab2} Calculated photoelectron polarizations for each atomic orbital based on Ref. \onlinecite{Trzhaskovskaya01} and corresponding to the incident photon energy 2000 eV. Values in parentheses are normalized photoelectron cross sections.
The final row reports the expected polarization of elastic photoelectrons over all atomic orbitals.
For simplicity, the solid angle of the MCP detector was not considered for the calculation.
Atomic orbitals with the binding energy larger than 600 eV are not listed due to the retarding field of the MCP without the sample bias voltage.
}
\begin{ruledtabular}
\begin{tabular}{cccc}
&Carbon&Silicon&Chromium\\
\hline
1$s_{1/2}$&0.99 (1)&&\\
2$s_{1/2}$&0.99 (5E-2)&1.00 (1)&\\
2$p_{1/2}$&0.16 (9E-4)&0.53 (2E-1)&0.76 (5E-1)\\
2$p_{3/2}$&&0.54 (3E-1)&0.76 (1)\\
3$s_{1/2}$&&1.00 (9E-2)&1.00 (1E-1)\\
3$p_{1/2}$&&0.52 (9E-3)&0.74 (6E-2)\\
3$p_{3/2}$&&&0.74 (1E-1)\\
3$d_{3/2}$&&&0.43 (5E-3)\\
3$d_{5/2}$&&&0.43 (7E-3)\\
4$s_{1/2}$&&&1.00 (7E-3)\\
\hline
\hline
Summarized Pol. &0.99&0.85&0.77\\
\end{tabular}
\end{ruledtabular}
\end{table}

Table \ref{tab2} lists the calculated photoelectron polarization, defined as ($J_{p}-J_{s}$)/($J_{p}+J_{s}$) and assuming linearly polarized incident photons at 2000 eV.
Orbitals that are unoccupied or have such large binding energies that their photoelectrons cannot reach the MCP at $h\nu$ = 2000 eV were excluded from the analysis.
The values in parentheses are the normalized photoionization cross sections.
In the case of carbon, photoelectrons originate predominantly from the $1s$ orbitals under these conditions ($\beta \simeq$ 2, $\delta$ and $\gamma$ are very small).
In contrast, the photoelectron polarizations for $p$ and $d$ orbitals deviate significantly from unity.
Their contributions are non-negligible for heavier elements such as Si and Cr.
Although the discussion above focuses only on elastic electrons, other isotropic components, such as Auger electrons, do not improve the contrast factor.
Therefore, light elements with fewer occupied orbitals should perform well as photon-polarization analyzers.
Because of its high availability and solid-phase stability at room temperature, carbon (graphite or glassy carbon) is a highly suitable target material for soft and tender X-ray polarimetry based on photoelectron angular distribution.

\section{Summary}
To summarize, we demonstrate herein that the angular distribution of photoelectrons, measured using a rotating-analyzer apparatus, serves as a tool for determining the linear polarization of incident X-ray beams.
The operational energy window spans from 1700 to 3000 eV with grounded carbon targets, and this range can be successfully extended down to 200 eV by applying a sample bias.
These results establish that, with a single experimental apparatus that does not require target replacement, the photoelectron angular distribution can serve as a robust probe for measuring the linear polarization of soft and tender X-rays across a wide energy range.


\begin{acknowledgments}
This work was supported by JSPS KAKENHI (Grant No. JP25K22223).
Some of the experiments were performed at NanoTerasu BL13U with the approval of the Japan Synchrotron Radiation Research Institute (Proposals No. 2025A9016 and No. 2025B9059).
We acknowledge fruitful discussion with T. Kinoshita and M. Suzuki that helped us understand the origin of dichromatic MCP signals.
The preparation of high-quality thin-film samples by T. Hatano is also acknowledged.
\end{acknowledgments}

\section*{Data Availability Statement}
The data that support the findings of this study are available within the article, its figures, and tables.

%

\end{document}